# 3/2 Fractional Quantum Hall Plateau in Confined Two-Dimensional Electron Gas

Hailong Fu[1,2,+], Yijia Wu[1,+], Ruoxi Zhang[1], Jian Sun[1], Pujia Shan[1], Pengjie Wang[1], Zheyi Zhu[1], L. N. Pfeiffer[3], K. W. West[3], Haiwen Liu[4], X. C. Xie[1,5,6], Xi Lin[1,5,6,*]

1 International Center for Quantum Materials, Peking University, Beijing 100871, China

2 Department of Physics, The Pennsylvania State University, University Park, Pennsylvania 16802, USA

3 Department of Electrical Engineering, Princeton University, Princeton, New Jersey 08544, USA

4 Center for Advanced Quantum Studies, Department of Physics, Beijing Normal University, Beijing 100875, China

5 Beijing Academy of Quantum Information Sciences, Beijing 100193, China

6 CAS Center for Excellence in Topological Quantum Computation, University of Chinese Academy of Sciences, Beijing 100190, China

+ HF and YW contribute equally to this work

* XL: xilin@pku.edu.cn

**Even-denominator fractional quantum Hall (FQH) states, such as 5/2 and 7/2, have been well known in a two-dimensional electron gas (2DEG) for decades and are still investigated as candidates of non-Abelian statistics. In this paper, we present the observation of a 3/2 FQH plateau in a single-layer 2DEG with lateral confinement at a bulk filling factor of 5/3. The 3/2 FQH plateau is quantized at $\left(\frac{h}{e^2}\right)/\left(\frac{3}{2}\right)$ within 0.02%, and can survive up to 300 mK. This even-denominator FQH plateau may imply intriguing edge structure and excitation in FQH system with lateral confinement. The observations in this work demonstrate that understanding the effect of the lateral confinement on the many-body system is critical in the pursuit of important theoretical proposals involving edge physics, such as the demonstration of non-Abelian statistics and the realization of braiding for fault-tolerant quantum computation.**

## Introduction

The FQH effect has been investigated to extend our understanding of strongly interacting particles in two dimensions since its discovery[1, 2]. Most of the observed fractional states are odd-denominator states[3] and can be explained under the framework of the composite fermion theory[4]. For example, the 1/3 FQH state can be mapped into the filling factor $\nu$ = 1 integer quantum Hall (IQH) state of composite fermions carrying fractional charge of e/3, which leads to a fractional $\frac{1}{3}\frac{e^2}{h}$ Hall conductance. At half-filled $N$ = 0 Landau level ($\nu$ = 1/2, 3/2), the absence of FQH states is attributed to the formation of Fermi sea of composite fermions[4, 5]. The discovery of the first even-denominator state, the 5/2 FQH state[6, 7], viewed as a paired state of composite fermions with e/4 fractional charge[8], further enriched the knowledge of interacting electrons. Recent studies such as phase transition to the stripe phase[9], and even-denominator states in other two-dimensional systems[10, 11], keep providing interesting physics at even-denominator filling factors.

The excitations in the FQH system carry fractional charge owing to the strong correlation between electrons. Besides, there is another equally famous way of charge fractionalization through topological soliton and topological phase transition[12, 13, 14], which is related to the boundary of a realistic experimental system. Theoretical treatments for the FQH effect usually consider an infinite 2DEG, but experiments can only measure finite samples with boundaries. The quantum Hall edge originating from the boundary was discussed[15] and later described by chiral Luttinger liquids[16]. Experiments based on the FQH edge physics have provided many exciting results and led to more efforts, such as the measurement of fractional charge[17], weak quasi-particle tunneling between edge currents[18, 19] and the search of the neutral modes[20]. Theoretically, a 2DEG confined in an interferometer is proposed to test the non-Abelian statistics directly[21], and the interference of the edge current has been measured[22]. Although the realistic boundary of a many-body system may complicate the theoretical treatment[23], the confinement can be a powerful experimental approach to manipulate the system. Even though the FQH effect has been studied for decades, systems of interacting electrons with confinement are still an open area for exploration.

Here we show an even-denominator FQH plateau quantized at $\left(\frac{h}{e^2}\right)/\left(\frac{3}{2}\right)$ under lateral confinement in a single-layer 2DEG. This plateau develops below 300 mK with a quantization of 0.02%. The current transmitting through the confined region also causes a quantized two-terminal conductance at $\frac{3}{2}\frac{e^2}{h}$, and a backscattered current induces a quantized tunneling conductance at $\frac{1}{6}\frac{e^2}{h}$.

## Results

### Gate-defined confinement

The Hall bar samples were made from a wafer of GaAs/AlGaAs heterostructures. The 2DEG is about 200 nm below the sample surface. Two separate gates were deposited on the surface for the confined region. A sketch of the Hall bar and measurement setup are shown in Fig. 1a. Fig. 1b shows the gate voltage dependence of the diagonal resistance $R_D$ across the confined region at zero magnetic field and $T = 6$ K. $R_D$ increases quickly from near zero to about 140 Ω when the gate voltage is swept from -1.0 V to -1.3 V, and below -1.3 V $R_D$ increases slowly, indicating the electrons underneath the gates are depleted when the gate voltage is not positive than -1.3 V. By estimating the transverse modes through the confined region, 140 Ω corresponds to a lateral confinement of ~2.1 μm, close to the lithography length. In order to achieve the confined region, $V_{gate} = -4.5$ V was applied to the gates to deplete the electrons underneath and around the gates. This gate voltage was kept for more than 6 hours above 4 K and during the cooling to 18 mK. This procedure named gate annealing is similar to that used in quasi-particle tunneling experiments for uniform density[18, 19, 24, 25]. At 18 mK, we kept gate voltage at -4.5 V, and measured the Hall traces between filling factor 1 and 2. Both the Hall resistance $R_{XY}$ in the bulk of the Hall bar and the diagonal resistance $R_D$ develop two IQH plateaus and several FQH plateaus, which indicates that well-defined compressible edge states and incompressible bulk states appear in the confined region, and the density inside the confined region and outside the confined region is nearly uniform (Fig. 1c).

**Four-terminal resistance**

In our samples, the confined region is formed and controlled by the gate voltages. Changing the gate voltage will tune the lateral confinement but not influence the bulk properties. As shown in Fig. 2a, when the gate voltage is changed to -1.3 V, the Hall resistance $R_{XY}$ does not change. Unexpectedly, the diagonal resistance $R_D$ develops an even-denominator FQH plateau quantized at $\left(\frac{h}{e^2}\right)/\left(\frac{3}{2}\right)$, with a quantization of 0.02%, the resolution of this experiment. The 3/2 FQH plateau of $R_D$ and the 5/3 FQH plateau of $R_{XY}$ center at almost the same magnetic field, and the width of these two plateaus are similar at 18 mK. Analogous to conventional FQH states, the width of the 3/2 plateau decreases with increasing temperature. Interestingly, the 3/2 FQH plateau is even wider than that of the 5/3 FQH plateau at 100 mK (Fig. 2b), and the 3/2 FQH plateau can survive up to 300 mK (Supplementary Figure 1), which indicates that the 3/2 plateau is more stable than all the previously observed even-denominator FQH plateaus in a single-layer 2DEG. The observation of such a 3/2 FQH plateau has been reproduced in different samples with different device sizes (see Supplementary Note 1).

The critical parameter to trigger the appearance of the 3/2 FQH plateau is the gate voltage. The diagonal resistance traces with different gate voltages at 18 mK are shown in Fig. 3a. The diagonal resistance develops the 5/3 FQH plateau with $V_{gate} = -2.8$ V, and becomes larger with less negative gate voltage. When the gate voltage is less negative than or equal to -2.0 V, the 3/2 FQH plateau appears. The gate voltage dependence of the 3/2 FQH plateau can also be demonstrated by the relationship between the gate voltage and the diagonal resistance. As shown in Fig. 3b, the diagonal resistance changes from $\left(\frac{h}{e^2}\right)/\left(\frac{3}{2}\right)$ to $\left(\frac{h}{e^2}\right)/\left(\frac{5}{3}\right)$ when the gate voltage is swept from -1.3 V to -3.0 V at the magnetic field of 7.67 T, near the center of the 5/3 FQH plateau in the bulk. At the same time, the Hall resistance is consistently quantized at $\left(\frac{h}{e^2}\right)/\left(\frac{5}{3}\right)$, indicating the 5/3 FQH state in the bulk of the Hall bar is not influenced by the gate voltage.

Another essential condition for the observation of the 3/2 FQH plateau is the gate annealing procedure with appropriate negative gate voltage. In our samples, $V_{gate}$ = -1.3 V is sufficient for the depletion of the electrons underneath the gates, as shown in Fig. 1b and Fig. 3c. During the measurements for the 3/2 FQH plateaus in Fig. 2, Fig. 3a and Fig. 3b, we did the annealing procedure with $V_{gate}$ = -4.5 V. If the gates are annealed at -1.5 V, even though enough to achieve the confined region, the 3/2 FQH plateau will not appear (Supplementary Figure 2).

In order to demonstrate the full depletion underneath the confinement gates, we fabricated a sample with a single top gate crossing the entire Hall bar. This gate was also annealed at $V_{gate}$ = -4.5 V, and the measurement was carried out at bulk filling factor 5/3 and $T$ = 18 mK. As shown in Fig. 3c, the conductance across the top gate is always zero when the gate voltage is more negative than -1.25 V. In another sample, we fabricated an individual gate (defined as arm gate in Fig. 3d) next to the confined region. As shown in Fig. 3d, a small negative gate voltage on this arm gate, far away from depleting values, is enough to break the 3/2 FQH plateau. These two control experiments demonstrated that the 3/2 FQH plateau is induced by the confined region.

In the confined region, a possible result caused by the annealing procedure and gate voltage change is quasi-particle tunneling. When $V_{gate}$ = -4.5 V is applied to the gates, the electrons both underneath and around the gates will be depleted. The annealing procedure may adjust the charge distribution of the donor layer at $T$ > 4 K, which causes a very sharp potential in the confined region at the measurement temperature range, and supports the 5/3 FQH state in the confined region (Fig. 1c). When the gate voltage is changed to less negative values, such as -2.0 V in Fig. 3a, at the lower temperature, the effective residual potential from both the gates and the donor layer may generate some disorders to the 2DEG, which facilitate the quasi-particle tunneling between the counter-propagating edge states in the confined region. More negative annealing gate voltage and less negative measurement gate voltage will generate more disorders, resulting in stronger quasi-particle tunneling. Strictly, the diagonal resistance is the sum of the Hall resistance in the bulk of the Hall bar and the tunneling resistance in the confined region. As a result, the variation of the diagonal resistance may be caused by the quasi-particle tunneling.

**Two-terminal conductance**

From the Landauer-Büttiker theory[26], when the diagonal resistance develops the 3/2 FQH plateau, the conductance $\frac{3}{2}\frac{e^2}{h}$ will transmit through the confined region, which has been verified in our measurements. Fig. 4a shows the two-terminal conductance traces at 18 mK. At filling factor 5/3, the conductance is quantized at $\frac{5}{3}\frac{e^2}{h}$ with $V_{gate}$ = -2.8 V and at $\frac{3}{2}\frac{e^2}{h}$ with $V_{gate}$ = -1.3 V. The conductance in the bulk of the Hall bar should be $\frac{5}{3}\frac{e^2}{h}$ at filling factor 5/3.

When the conductance transmitting the confined region is $\frac{3}{2}\frac{e^2}{h}$, the tunneling conductance will be $\frac{1}{6}\frac{e^2}{h}$. The corresponding propagating edge channels are sketched in Fig. 4b, where the dashed lines represent the tunneling conductance quantized at $\frac{1}{6}\frac{e^2}{h}$. This edge channel picture has also been verified by four-terminal resistance and two-terminal conductance measurements with the reversed magnetic field (Supplementary Figure 4 and Figure 5).

## Discussion

In general, the density of the gate-defined confinement region is smaller than that of the bulk, and affected by the gate voltage. However, the gate annealing procedure can keep uniform density for the whole sample in our measurements. This can be verified by the magnetic field dependence of the IQH and FQH plateaus in Hall traces in Fig. 1c and Supplementary Figure 6. By comparing the diagonal resistance and the Hall resistance as a function of magnetic field for the gating condition of developing the 3/2 FQH plateau (Fig. 3a), the 3/2 plateau in this measurement appears at the filling factor of 5/3. What's more, if the density of the confinement is reduced to filling factor 3/2 when the 3/2 plateau shows up at -1.8 V (Fig. 3a), then further negatively charging the top gate will only maintain or keep reducing the density in the confinement. As a result, the 5/3 plateau could not show up anymore at the filling factor 3/2 or at a smaller filling factor, which is contradicted to the experimental observation (-2.8 V at Fig. 3a). This argument serves as an additional clue that the 3/2 FQH plateau appears at the filling factor of 5/3. The 3/2 FQH plateau has also been observed in Charles M. Marcus group, shown in Yiming Zhang's thesis[27]. In their measurements, the 3/2 FQH plateaus appeared in samples with both uniform density and non-uniform density. From the corresponding *R*-*B* features between $R_D$ and $R_{XY}$ in Fig. 1c and Supplementary Figure 6, the 3/2 FQH plateau we observed is unlikely induced only by the density variation in the confined region.

Disorder can drive the collapse of energy gap and then induce phase transition from the FQH state to another state[28, 29]. At $\nu = 5/3$ FQH state, the edge state could be complicated due to potential edge reconstruction and neutral mode[20, 23, 30]. The formation of the 3/2 FQH plateau could be caused by a more complicated edge reconstruction with lateral confinement at $\nu = 5/3$. In our system, the disorders induced by the residual potential in the confined region could also help with the collapse of the 5/3 FQH state and the formation of intriguing edge structure and excitation with fractional charge. The 3/2 FQH plateau has been found in the single-layer two-dimensional system of ZnO where Landau level crossing induced by tilted magnetic field is needed[10], while there is no in-plane magnetic field in our measurements. The exact origin of the 3/2 FQH plateau remains an open theoretical question to be investigated.

## Methods

### Sample fabrication

The samples were made from a wafer of GaAs/AlGaAs heterostructures. The center of the 30 nm quantum well is 210 nm below the surface, with modulated Si doping at ~100 nm below and above the quantum well center. The mobility is higher than $1.0\times10^7$ $cm^2$ $V^{-1}$ $s^{-1}$ at 20 mK. A 150 μm wide Hall bar was shaped by wet etching. Ohmic contacts were made of Pt/Au/Ge alloy deposited by e-beam evaporation and annealed at 550 °C for 100 s. Two separate Cr/Au gates were deposited on the surface of the sample to form the confined region. In this experiment, samples with different confined region sizes were fabricated and measured. The dimensions of these geometries are $1\times2$ $\mu m^2$, $0.3\times2$ $\mu m^2$ and $1.5\times3$ $\mu m^2$ respectively. The confined region on the samples used in the main text is the size of $1\times2$ $\mu m^2$. The data from other samples are shown in the Supplementary Figures 1, 2 and 3.

## Measurement techniques

The measurements were carried out in a dilution refrigerator with a base refrigerator temperature below 6 mK. The temperatures given in this work are electron temperatures. In the four-terminal resistance measurements, the Hall resistance $R_{XY}$ and the diagonal resistance $R_D$ were measured simultaneously using lock-in techniques at 6.47 Hz with 1 nA. The two-terminal conductance measurements were also carried out using lock-in techniques at 6.47 Hz with voltage excitations smaller than 35 μV.

## Data availability

The authors declare that the main data supporting the findings of this study are available within the article and its Supplementary Materials. Extra data are available from the corresponding author upon request. The source data underlying Figs 1b, 1c, 2, 3, 4a and Supplementary Figs 1- 6 are provided as a Source Data file.

**Acknowledgements**

We thank Rui-Rui Du, Jainendra K. Jain, Hua Jiang, Jie Liu, Yang Liu, Junren Shi, Xin Wan, Ding Zhang, Jun Zhu and Junyi Zhang for discussions. We thank Charles M. Marcus for reminding us of his observation on the 3/2 FQH plateau in Yiming Zhang's thesis. The work is supported by the NBRPC (2015CB921100), NSFC (11674009), National Key Research and Development Program of China (2017YFA0303301), NSFC (11534001), the Beijing Natural Science Foundation (JQ18002), the Strategic Priority Research Program of Chinese Academy of Sciences (XDB28000000) and the Fundamental Research Funds for the Central Universities (11534001). H. F. acknowledges the support of the Eberly Postdoc Fellowship at Penn State University. The work at Princeton University was funded by the Gordon and Betty Moore Foundation through the EPiQS initiative Grant GBMF4420, by the National Science Foundation MRSEC Grant DMR-1420541, and by the Keck Foundation.

**Author Contributions**

H.F., R.Z., J.S., P.S., and P.W. performed the measurements. H.F. and R.Z. fabricated the devices. L.N.P. and K.W.W. prepared and supplied the GaAs wafer. H.F. and X.L. initiated the research. H.F., R.Z., J.S., Z.Z., P.W. and X.L. analyzed the results. Y.W., H.L. and X.C.X. provided the theoretical attempt. H.F., Y.W., H.L., X.C.X. and X.L. discussed the results and wrote the manuscript.

**Competing interests**

The authors declare no competing interests.

## Figure legends

**Figure 1 | Sample information and the diagonal resistance and the Hall resistance traces. a,** A sketch of the Hall bar and the measurement setup. **b,** Gate voltage dependence of the diagonal resistance $R_D$ at B = 0 T and $T$ = 6 K. The inset is an SEM picture of a device with the same gate geometry as that used in this experiment. The lithography dimension is 1×2 μm$^2$. **c,** The diagonal resistance and the Hall resistance traces with $V_{gate}$ = -4.5 V at 18 mK. The black line is the Hall resistance $R_{XY}$ and the blue line is the diagonal resistance $R_D$. Both of them develop a series of IQH and FQH plateaus. $R_D$ was measured from contact 3 to contact 4, and $R_{XY}$ was from contact 6 to contact 5. Source data are provided as a Source Data file.

**Figure 2 | The diagonal resistance and the Hall resistance traces around the $\nu$ = 5/3 FQH state with $V_{gate}$ = -1.3 V. a,** At 18 mK, the Hall resistance (the black line) is quantized at $\left(\frac{h}{e^2}\right)/\left(\frac{5}{3}\right)$, and the diagonal resistance (the red line) is quantized at $\left(\frac{h}{e^2}\right)/\left(\frac{3}{2}\right)$. **b,** Temperature dependence of the 3/2 FQH plateau of the diagonal resistance and the 5/3 FQH plateau of the Hall resistance in the bulk of the Hall bar. Source data are provided as a Source Data file.

**Figure 3 | Gate voltage dependence of the diagonal resistance, the Hall resistance and the two-terminal conductance. a,** Magnetic field dependence of the diagonal resistance with different gate voltages. **b,** Gate voltage dependence of the diagonal resistance and the Hall resistance at 7.67 T. The gate voltage was changed from -1.3 V to -3.0 V. **c,** Gate voltage dependence of the two-terminal conductance across the single top gate. The gate was annealed at -4.5 V. The conductance was measured at bulk filling factor 5/3. The measurement was carried out by applying a voltage excitation to the source contact and measuring the current from drain contact at the other side of the mesa. The inset is the sketch of the device and the device was made from the same wafer as that used in **a**. The minimum width of the top gate is 1.5 μm, the same width as that of the gates used in one sample, with which the 3/2 plateau is observed (Supplementary Figure 2). **d,** Arm gate voltage ($V_{arm}$) dependence of the diagonal resistance at bulk filling factor 5/3. The confinement gates were annealed at -4.5 V. The arm gate was not annealed. During the measurement the confinement gates were kept at -1.3 V. The inset is the sketch of the device and the device was made from the same wafer as that used in **a**. The dimension of the confinement is 1×2 μm$^2$. All these measurements were performed at 18 mK. Source data are provided as a Source Data file.

**Figure 4 | Gate voltage dependence of the two-terminal conductance and the sketch of the propagating edge channels. a,** Two-terminal conductance traces at 18 mK with $V_{gate}$ = -1.3 V and $V_{gate}$ = -2.8 V respectively. The measurement was carried out by applying a voltage excitation to the contact $S$ and measuring the current from contact $D$ (contacts $S$ and $D$ are the same configurations as that shown in Fig. 1a). **b,** The sketch of the propagating edge channels for the formation of the 3/2 FQH plateau. In the bulk, the filling factor is 5/3. In the confined region, the straight lines represent the transmitting conductance quantized at $\frac{3}{2}\frac{e^2}{h}$, and the dashed lines represent the tunneling conductance quantized at $\frac{1}{6}\frac{e^2}{h}$. Source data are provided as a Source Data file.

Figure 1

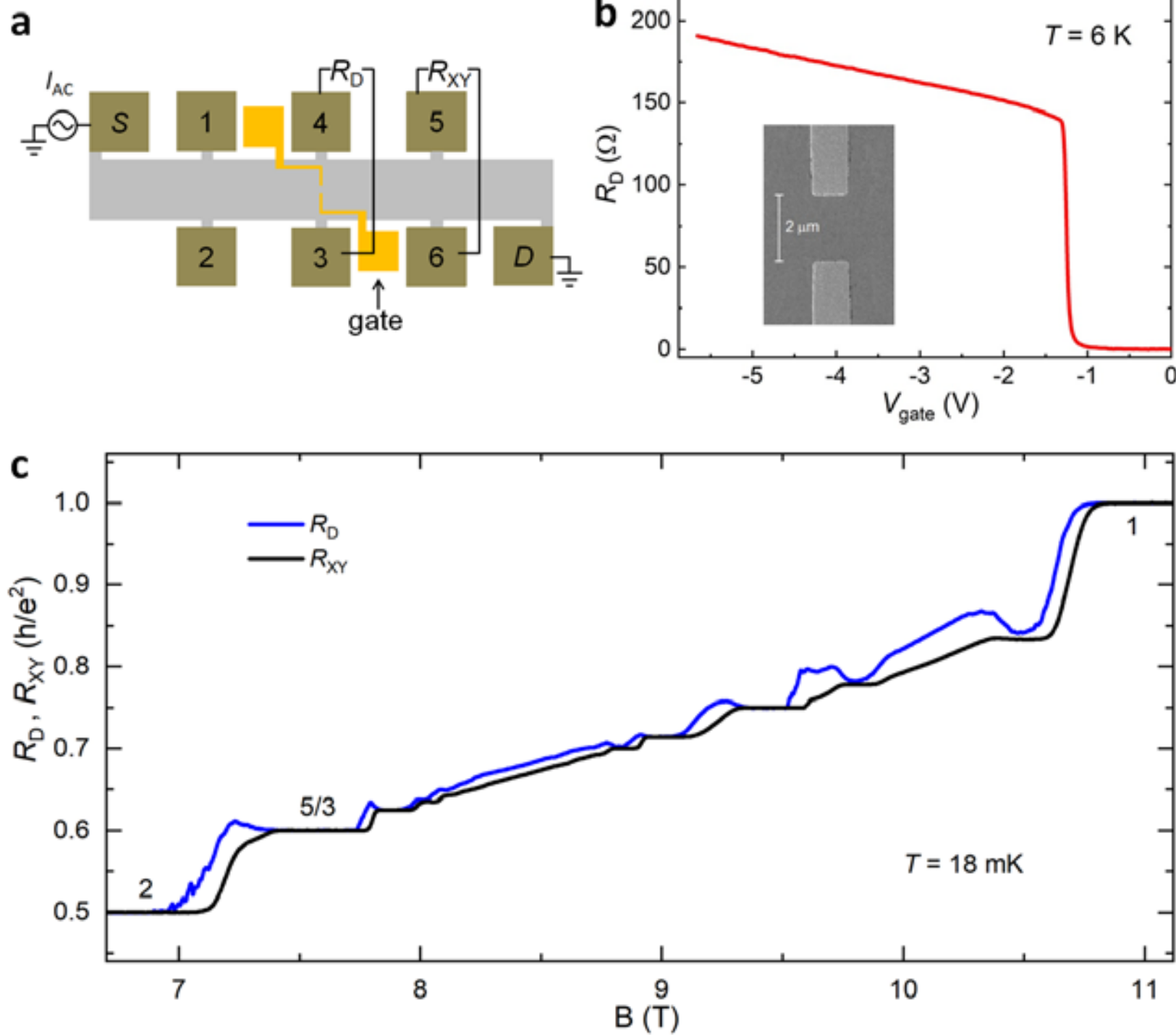

Figure 2

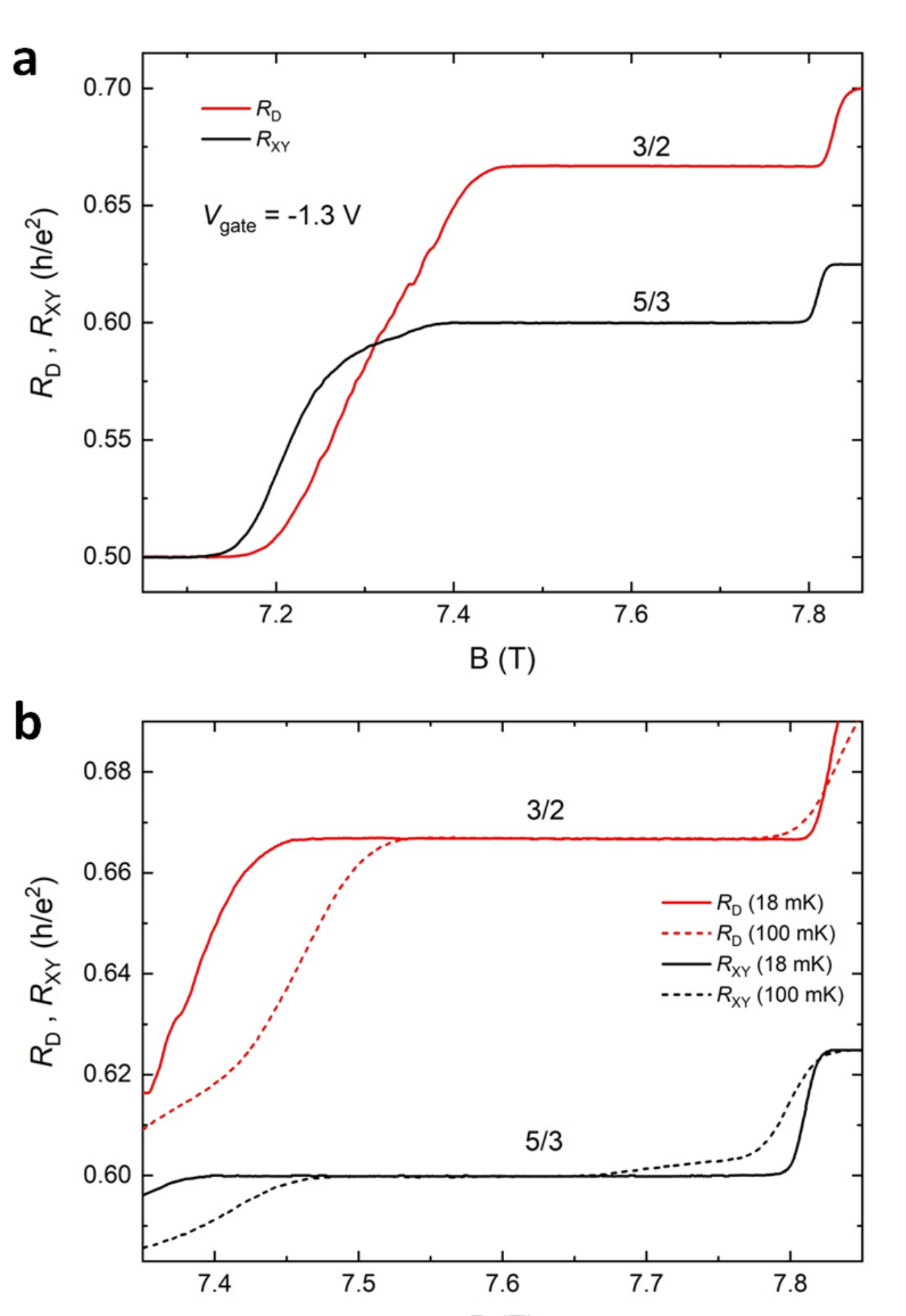

Figure 3

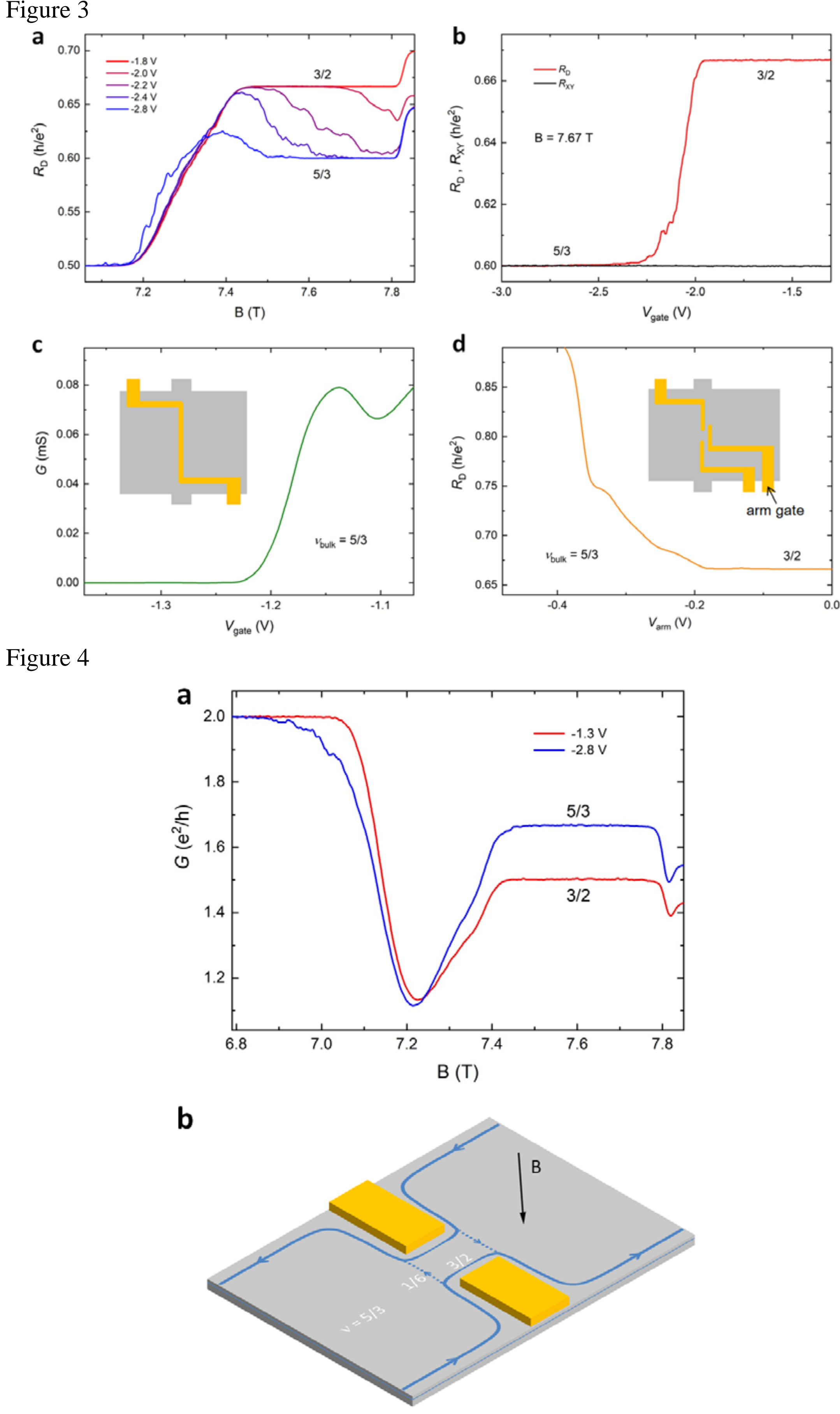

Supplementary Information

# "3/2 Fractional Quantum Hall Plateau in Confined Two-Dimensional Electron Gas"

Fu et al.

## Supplementary Note 1: Experimental results

This section shows more experimental results measured in different samples. Data in Supplementary Figure 1, Figure 2 and Figure 3 come from three more samples, and the appearance of the 3/2 FQH plateau is consistent with the results in the main text. Influence of the annealing procedure on the 3/2 FQH plateau is shown Supplementary Figure 2.

Supplementary Figure 4 and Figure 5 show the results measured in the opposite magnetic field directions. The diagonal resistance $R_{\mathrm{D}}$ in our experimental setup is determined by the voltage difference between contacts 3 and 4, explicitly as $R_{\mathrm{D}} = \frac{V_3 - V_4}{I_{\mathrm{AC}}}$ (Fig. 1 in the main text). Therefore, $R_{\mathrm{D}} = \frac{V_3 - V_6}{I_{\mathrm{AC}}} + \frac{V_6 - V_4}{I_{\mathrm{AC}}} = R_{\mathrm{L}} + R_{\mathrm{XY}}$; the observation that the diagonal resistance $R_{\mathrm{D}}$ is larger than the Hall resistance $R_{\mathrm{XY}}$ indicates that part of the edge channel does not pass through the confined region (Fig. 4b in the main text). If we keep the same measurement setup and reverse the directions of the edge channel by switching the magnetic field, we should measure $R_{\mathrm{L}}$-$|R_{\mathrm{XY}}|$. Now the sign of $R_{\mathrm{XY}}$ will be changed by the magnetic field while the sign of $R_{\mathrm{L}}$ will remain the same because the source and drain remain the same positions. Because $V_3$ is smaller than $V_4$ with a reversed magnetic field, the measured diagonal resistance $R_{\mathrm{D}}$ should be also negative. Experimentally, we did observe a new plateau quantized at negative $\left(\frac{h}{e^2}\right) / \left(\frac{15}{8}\right)$ with the same experimental setup and opposite magnetic field, as shown in Supplementary Figure 4. The average of resistance $\left(\frac{h}{e^2}\right) / \left(\frac{3}{2}\right)$ and resistance $\left(\frac{h}{e^2}\right) / \left(\frac{15}{8}\right)$ is $\left(\frac{h}{e^2}\right) / \left(\frac{5}{3}\right)$, which is identical to $R_{\mathrm{XY}}$, as expected in the picture of edge channel. The observation of the -15/8 FQH plateau verifies the edge channel reflection in the confined region.

Supplementary Figure 6 shows the Hall resistance and the diagonal resistance traces with different gate voltages to demonstrate the nearly uniform density in the whole sample. In general, more negative gate voltage will cause lower density in the confined region. As shown in Supplementary Figure 6a, if the gates have never been annealed, more negative gate voltages result in larger slopes of the diagonal resistance traces, corresponding to lower densities in the confined region. And the diagonal resistance traces with different densities develop with the same IQH plateaus at different magnetic fields. As shown in Supplementary Figure 6b, when the gates have been annealed with -4.5 V, the slopes of the diagonal resistance traces with less negative gate voltages are larger than that with -4.5 V, and no well-defined IQH plateaus appear, which may be caused by edge current backscattering rather than the density difference. If we extract the density from the slopes at -1.3 V from Supplementary Figure 6b between 0.15 T and 0.35 T, and calculate the filling factors at high magnetic field, then the $\nu = 2$ state unreasonably lies between filling factor 1.9 and 1.55, as shown in Supplementary Figure 6c. In addition, the 3/2 plateau doesn't match the filling factor 3/2 even from the density calculated from the low field data (red top x-axis). We believe the density in the whole sample is nearly uniform.

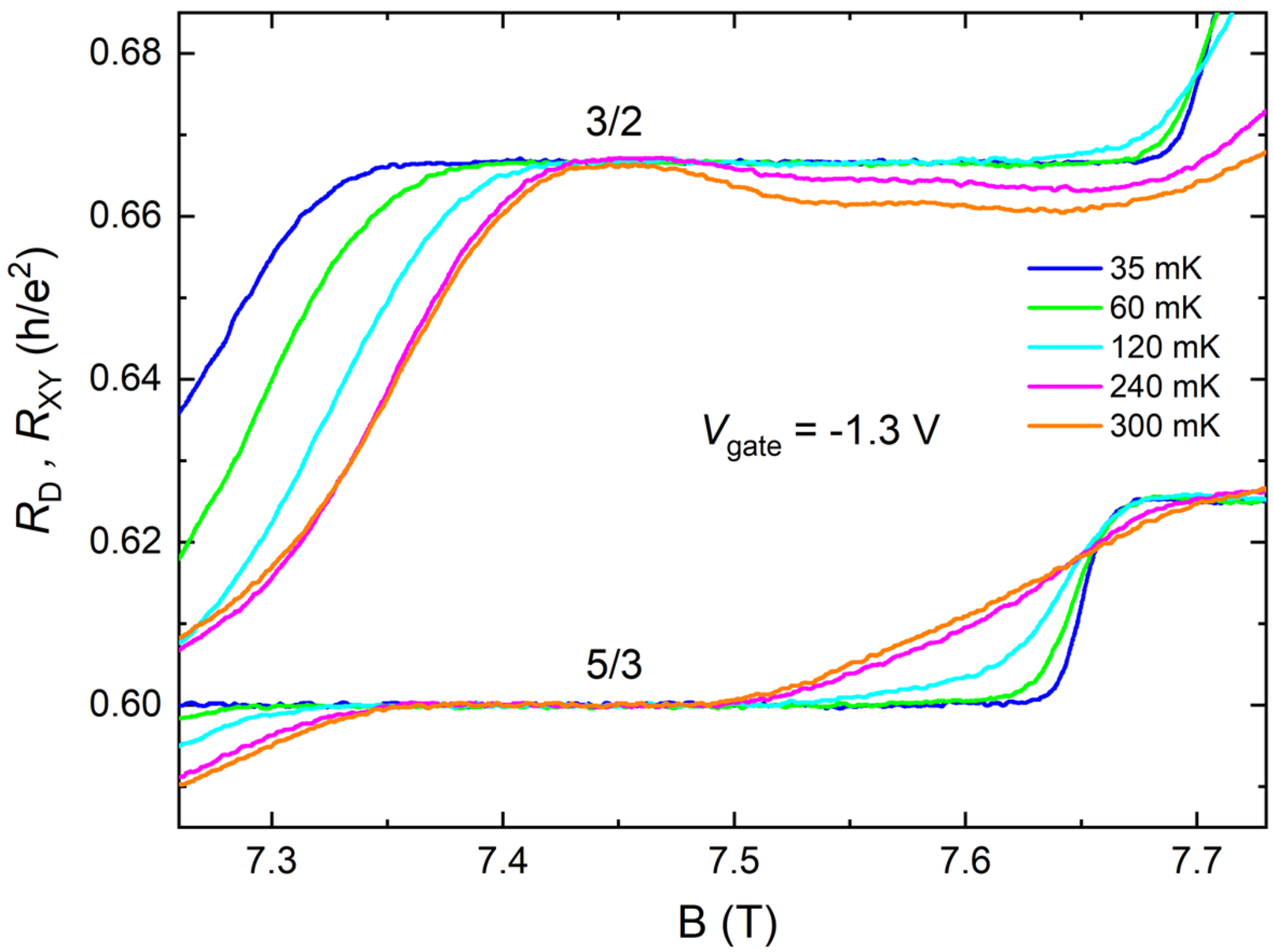

**Supplementary Figure 1. The diagonal resistance and the Hall resistance traces around the 5/3 FQH state with $V_{gate}$ = -1.3 V at different temperatures.** The widths of the 3/2 and 5/3 FQH plateaus decrease with increasing temperature, and the 3/2 FQH plateau can even survive up to 300 mK. Data are from another sample with the same confined region as that used in the main text. Source data are provided as a Source Data file.

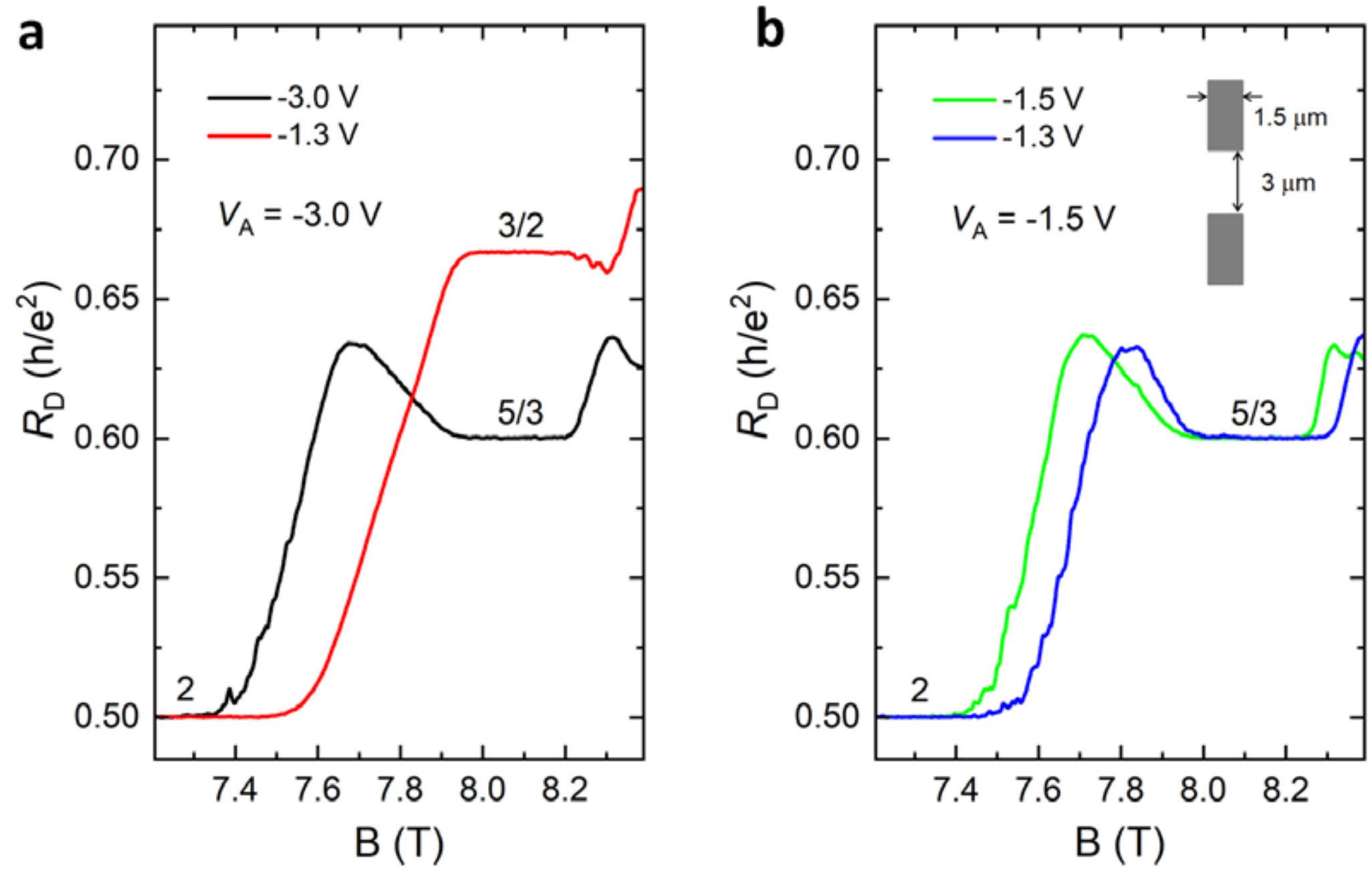

**Supplementary Figure 2. Influence of the annealing procedure on the 3/2 FQH plateau. a,** The gates were annealed at -3.0 V. The diagonal resistance develops the 5/3 FQH plateau with $V_{gate}$ = -3.0 V and the 3/2 FQH plateau with $V_{gate}$ = -1.3 V. **b,** The gates were annealed at -1.5 V. The diagonal resistance only develops the 5/3 FQH plateau with both $V_{gate}$ = -1.5 V and $V_{gate}$ = -1.3 V. The confined region of this device is shown in the inset of **b** as a sketch. Source data are provided as a Source Data file.

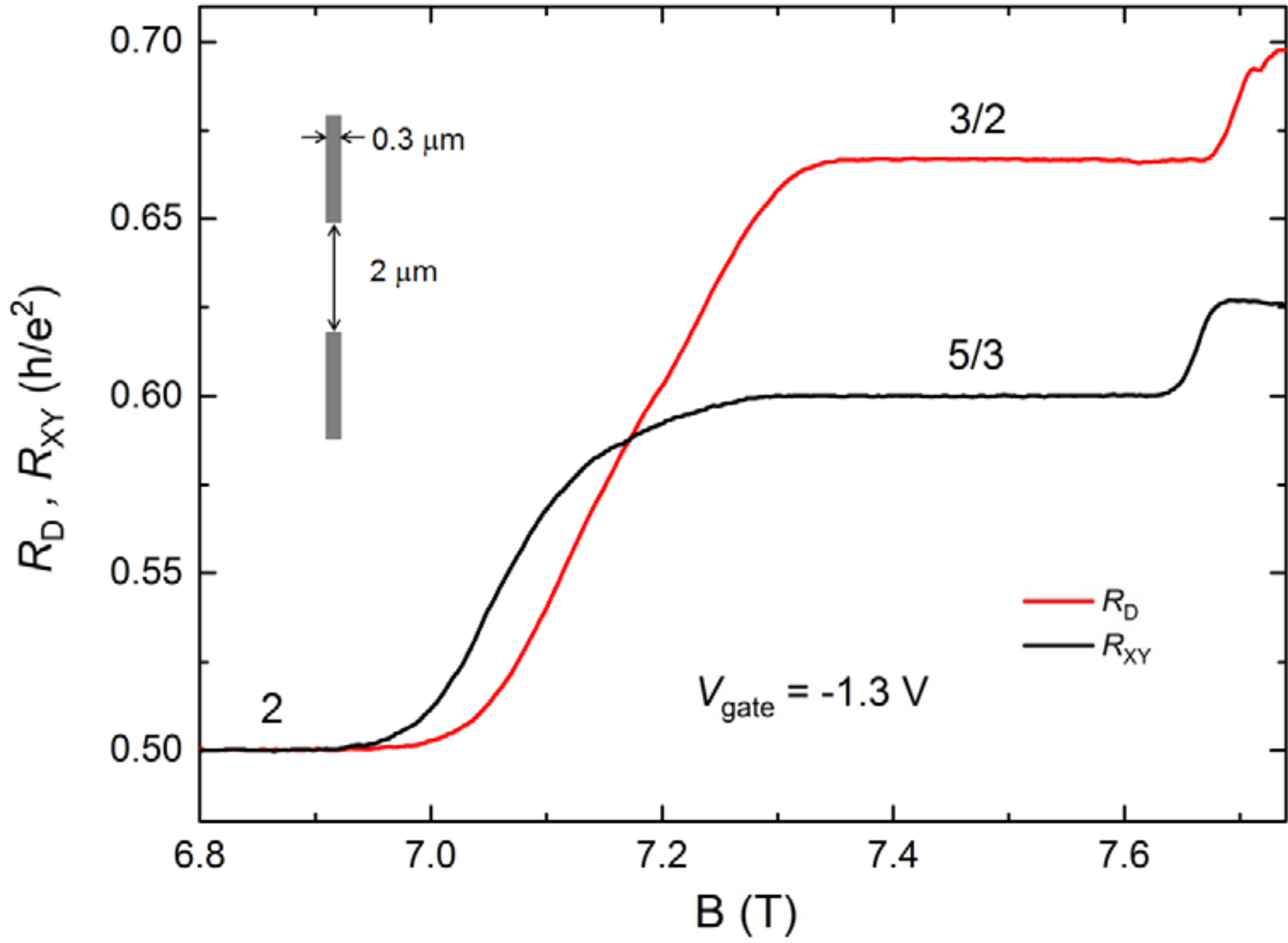

**Supplementary Figure 3. The diagonal resistance and the Hall resistance traces with $V_{gate}$ = -1.3 V at 35 mK from another device.** The gates were annealed at -4.5 V. The diagonal resistance develops the 3/2 FQH plateau and the Hall resistance develops the 5/3 FQH plateau. The inset is a sketch of the confined region. Source data are provided as a Source Data file.

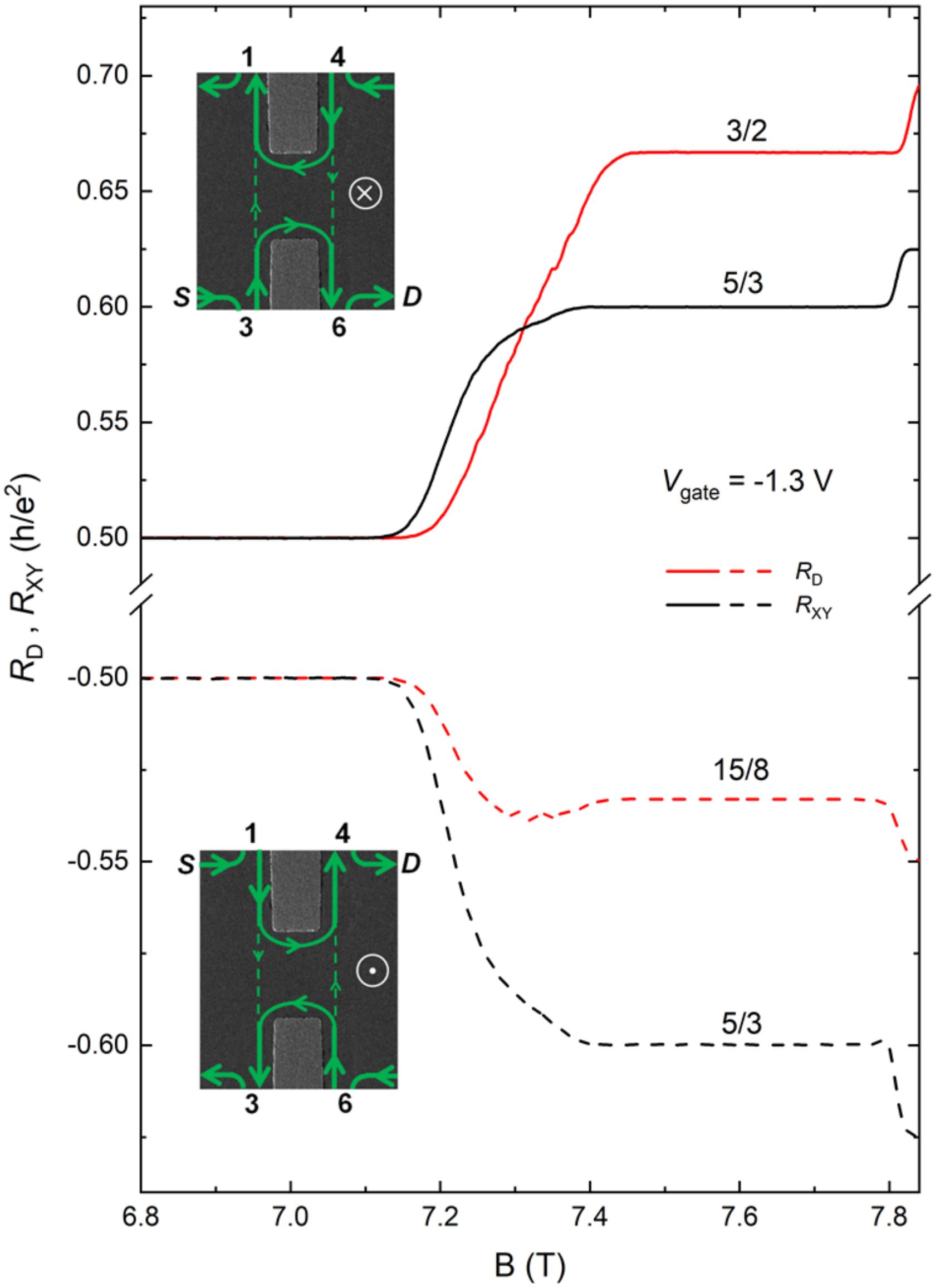

**Supplementary Figure 4. The diagonal resistance and the Hall resistance traces in opposite magnetic field directions with $V_{gate}$ = -1.3 V at 18 mK.** $R_D$ was measured from contact 3 to contact 4 and $R_{XY}$ was measured effectively as from contact 6 to contact 4. When the magnetic field direction is into the page (the upper inset), $R_{XY}$ is quantized at $\left(\frac{h}{e^2}\right)/\left(\frac{5}{3}\right)$ (solid black line) and $R_D$ is quantized at $\left(\frac{h}{e^2}\right)/\left(\frac{3}{2}\right)$ (solid red line). When the magnetic field direction is out of the page, only the propagating directions of the edge channels are reversed (the lower inset). As a result, $R_{XY}$ is quantized at $-\left(\frac{h}{e^2}\right)/\left(\frac{5}{3}\right)$ (dashed black line) and $R_D$ is quantized at $-\left(\frac{h}{e^2}\right)/\left(\frac{15}{8}\right)$ (dashed red line). Data are from the same sample used in the main text. Source data are provided as a Source Data file.

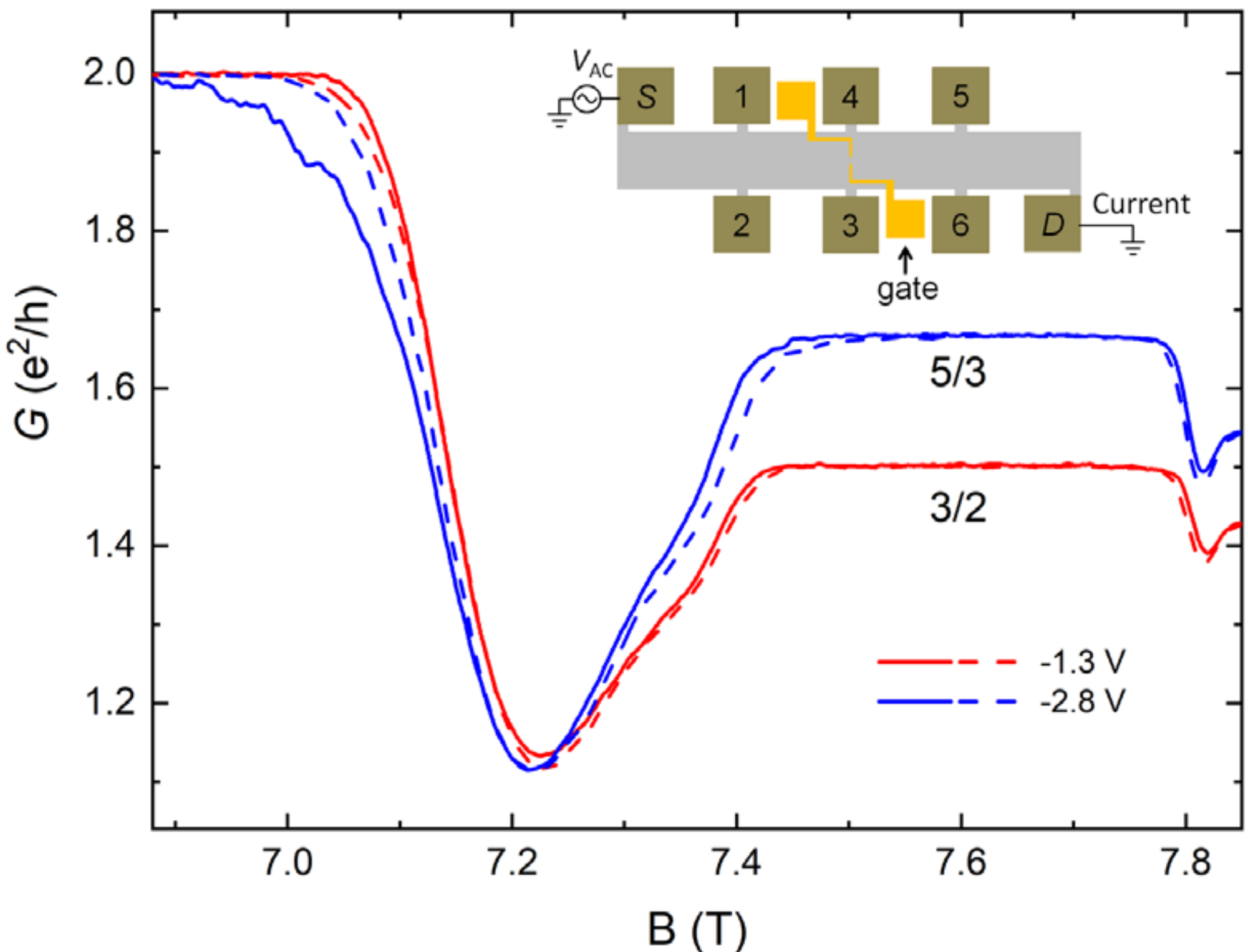

**Supplementary Figure 5. Two-terminal conductance traces in opposite magnetic field directions at 18 mK.** The conductance was obtained by applying a voltage excitation to the contact *S* and measuring the current from contact *D*. Solid lines and dashed lines represent data measured in the opposite magnetic field directions respectively. The inset is the sketch of the Hall bar and the measurement setup. The observation of the 3/2 FQH conductance plateau in both field directions verifies the edge channel reflection picture in the confined region. Data are from the same sample used in the main text. Source data are provided as a Source Data file.

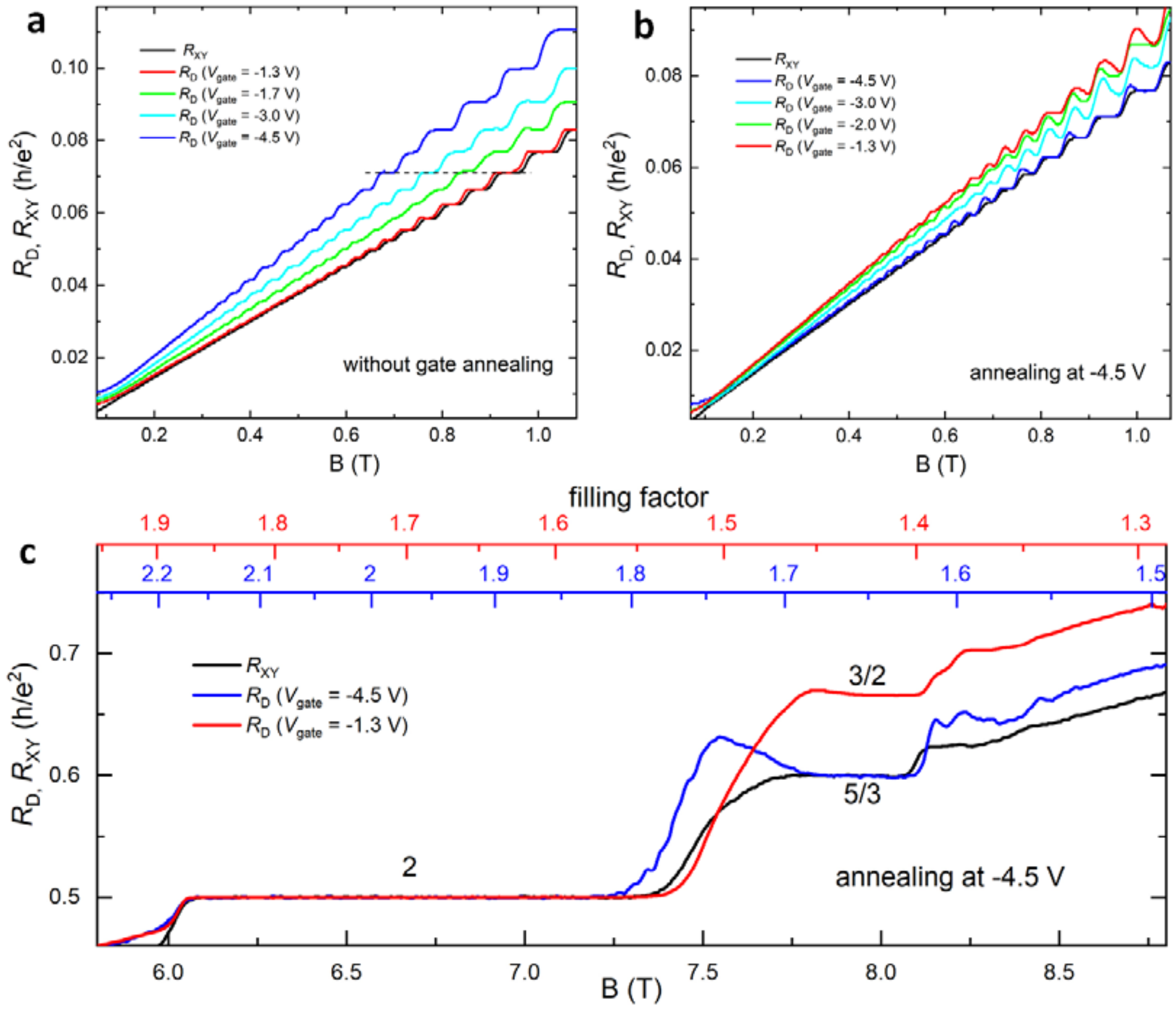

**Supplementary Figure 6. The diagonal resistance and the Hall resistance traces with different gate voltages. a,** The diagonal resistances and the Hall resistance traces at low magnetic field without gate annealing. The horizontal dashed line is an eye-guide to show that the same IQH plateau appears with different electron densities. **b,** The diagonal resistances and the Hall resistance traces at low magnetic field with gate annealing at -4.5 V. **c,** The diagonal resistances and the Hall resistance traces at high magnetic field with gate annealing at -4.5 V. Top red (blue) axis shows the filling factors calculated from the diagonal resistance traces with -1.3 V (-4.5 V) at low magnetic field from **b**. The difference in *R*-B traces at low field should result from confinement effect but not the density difference, because the $\nu = 2$ state at $V_{gate}$ = -1.3 V should not lie between filling factors 1.9 and 1.55. Source data are provided as a Source Data file.

## Supplementary Note 2: Preliminary theoretical attempt

A tentative explanation to the origin of the 3/2 FQH plateau could be a further 1/2 quantization of the quasi-particles with e/3 fractional charge in the 5/3 FQH state, which causes a $\frac{1}{6}\frac{e^2}{h}$ backscattering and therefore the 3/2 quantum Hall plateau. Such 1/2 quantization is proposed to be related to the half-charged Jackiw-Rebbi zero-mode [1] induced by a topological phase transition near the confined geometry.

A heuristic description could be that, in the perspective of composite fermion, as an illustration, the edge channels of the $\nu = 5/3$ FQH state can be treated as containing two downstream $e^* = e/3$ charged edge channels with different chemical potentials, moving in the presence of the residual magnetic field. A one-dimensional four-band Hamiltonian [2] can be constructed in the basis of these $e^* = e/3$ edge channels as $(R_+, L_-, R_-, L_+)$ (where R/L for right-/left-moving channel in the opposite edges, and $\pm$ for edge channels with higher/lower chemical potential, respectively) as $H = H_0 + \Delta H$. Here $H_0 = \hat{p}_x \tau_0 \rho_3 + t\tau_1\rho_1 + \mu\tau_3\rho_3$ includes the linearized kinetic energy term $\hat{p}_x$, inter-edge tunneling term $t$, and chemical potential $\mu$ [$\tau_0$ is identity matrix and $\tau_{i=1,2,3}$ are Pauli matrices acting on the $(R_+/L_-, R_-/L_+)^T$ spinor, and ρ-matrices are Pauli matrices working in the right-/left-moving space], while $\Delta H = \Delta \cdot \tau_0 \rho_1$ is a perturbative ($\Delta$ is small but non-zero) coupling term between channels with different chemical potentials.

The dispersions for the unperturbed four-band Hamiltonian $H_0$ are: $\pm\mu \pm \sqrt{k_x^2 + t^2}$. For non-zero $t$, the uppermost band $|\Psi_u\rangle = C\left(1,0,0,-k_x/t + \sqrt{k_x^2+t^2}/t\right)^T$ and the lowermost band $|\Psi_l\rangle = C\left(0,1,k_x/t - \sqrt{k_x^2+t^2}/t\,,0\right)^T$ ($C$ is the normalization constant) are always gapped (see Supplementary Figure 7), therefore they are not involved in the topological phase transition and contribute $\frac{1}{3}\frac{e^2}{h}$ Hall conductance as ever. However, for the intermediate two bands

$E_\pm = \mp\mu \pm \sqrt{k_x^2 + t^2}$ with $|\Psi_-\rangle = C\left(k_x/t - \sqrt{k_x^2+t^2}/t\,,0,0,1\right)^T$ and $|\Psi_+\rangle = C\left(0,-k_x/t + \sqrt{k_x^2+t^2}/t,1,0\right)^T$, projecting $\Delta H$ onto $(|\Psi_-\rangle, |\Psi_+\rangle)^T$ leads to an effective Hamiltonian equivalent to one-dimensional topological insulator:

$$H_{\text{eff}} = \Delta \frac{k_x}{\sqrt{k_x^2+t^2}}\sigma_x + \left(\mu - \sqrt{k_x^2+t^2}\right)\sigma_z \tag{1}$$

where the $\sigma$-matrices are Pauli matrices acting on the higher-/lower-chemical potential spinor. The competition between the inter-edge tunneling strength $t$ and the chemical potential $\mu$ leads to a topological phase transition (see Supplementary Figure 7). At the domain wall separating the tunneling-dominated region ($t > \mu$; near the confined geometry) and the magnetic-field-dominated region ($t < \mu$; far away from the confined geometry), there are topologically protected Jackiw-Rebbi type bound states possessing $e^*/2 = e/6$ charge [1, 2].

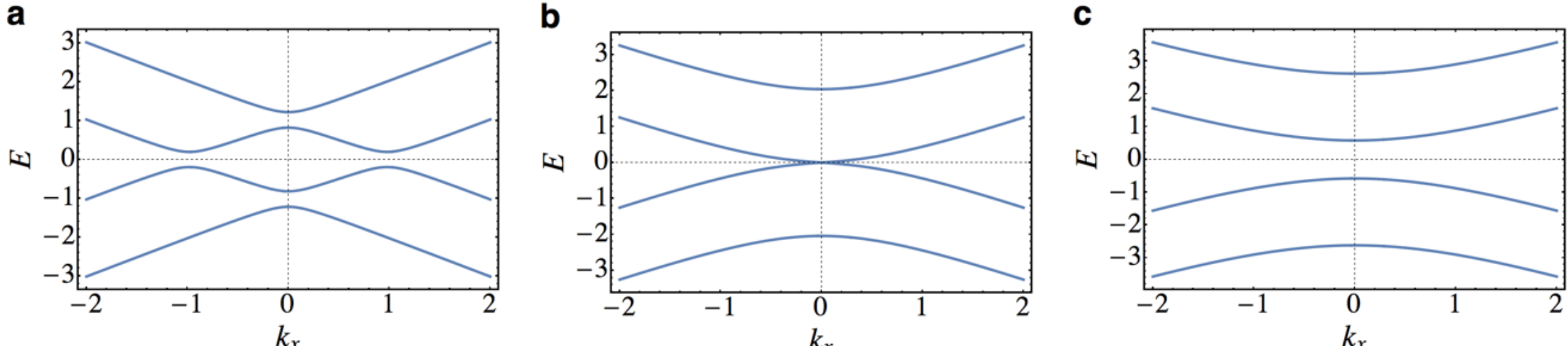

**Supplementary Figure 7. Dispersions and topological phase transition of the one-dimensional four-band Hamiltonian $H = H_0 + \Delta H$. a,** Magnetic-field-dominated region ($t < \mu$). $\mu = 1, \Delta = 0.2$, and $t = 0.2$. **b,** Phase transition point ($t \approx \mu$). $\mu = 1, \Delta = 0.2$, and $t = 1.02$. **c,** Tunneling-dominated region ($t > \mu$). $\mu = 1, \Delta = 0.2$, and $t = 1.6$. The uppermost and the lowermost bands are always gapped, while the intermediate two bands can be approximately treated as a one-dimensional topological insulator. The Jackiw-Rebbi type zero-modes are bounded at the domain wall separating these two topologically distinct phases (**a** and **c**).

However, compared with the two-dimensional geometry of the experimental setup, the model described above is actually a quasi-one-dimensional one. There could be several ways to generalize the Jackiw-Rebbi zero-mode possessing half charge $e^*/2$ to the two-dimensional case. Two-dimensional $e^*/2$ zero-mode could be realized by introducing a vortex [3, 4]. However, such an $e^*/2$ zero-mode is bound state uninvolved in the electrical transport except a charge pump is provided [5]. Another proposed generalization could be the Moore-Read state [6, 7]. Here, in this two-dimensional electron system, we suppose that a promising two-dimensional generalization of the Jackiw-Rebbi type zero-mode could be the massless Dirac fermion, which is analogous to the generalization from the localized end-estate (in one-dimensional topological insulator) to the edge states with linear dispersion (in two-dimensional topological insulator). Such generalization is also consistent with Son's theory [8] that the composite fermions in the half-filled Landau level with particle-hole symmetry are massless Dirac fermions. In the presence of the residual magnetic field, the massless Dirac fermions' zeroth Landau level forms edge channel carrying half quantized Hall conductance $\frac{1}{2} \cdot \frac{1}{3}\frac{e^2}{h}$ [9], which is analogous to the half-charged bound state in the one-dimensional case. Consequently, such an edge channel with $\frac{1}{6}\frac{e^2}{h}$ conductance in the tunneling-dominated region, may accounts for the backscattering conductance and therefore explains the emergence of the $\frac{3}{2}\frac{e^2}{h}$ Hall plateau.

Such a heuristic model is stated here as an illustration of the proposed theoretical scheme. The actual edge states of the 5/3 FQH state could be much more complicated due to edge reconstruction, such as containing two downstream $e^* = e/3$ charged edge channels as well as two upstream neutral charge modes [10, 11]. Generally, the inter-edge tunneling strengths are different for different edge channels, which are relevant to the spatial distribution of the edge channels' wave-functions. Besides, in realistic experimental device, the inter-edge tunneling strength could be spatially inhomogeneous, so that the tunneling may also transfer quasiparticles between opposite edges and therefore induce a position dependence of the chemical potential. Moreover, the free fermion description of the fractionally-charged edge channels is also inaccurate. For these reasons and the different possibilities of the generalization of Jackiw-Rebbi zero-mode mentioned above, the theoretical descriptions here are provided as a tentative hypothesis. A solid proof of the origin of such a novel 3/2 FQH state still remains as an open question to be further investigated.

**Supplementary references**